\documentstyle[epsf,epsfig,here,12pt]{article}
\def\journal#1&#2(#3){\unskip, \sl #1\unskip~\bf\ignorespaces #2\rm
(19#3)}

\def\beq{\begin{equation}}
\def\eeq{\end{equation}}
\def\bea{\begin{eqnarray}}
\def\eea{\end{eqnarray}}

\newcommand{\Du}{\hbox{$\Delta u$}}
\newcommand{\Dd}{\hbox{$\Delta d$}}
\newcommand{\Ds}{\hbox{$\Delta s$}}

\def\ZZ{\raise0.12em\hbox{\scriptsize{$
\not
\kern0.15em\not
\kern-0.21em\lower0.2em
\vbox{\hrule width 0.52em height 0.06em depth 0pt}
\kern-0.50em\raise0.65em
\vbox{\hrule width 0.52em height 0.06em depth 0pt}
\,$}}}

\catcode`\@=11 
\def\lsim{\mathrel{\mathpalette\@versim<}}
\def\gsim{\mathrel{\mathpalette\@versim>}}
\def\@versim#1#2{\vcenter{\offinterlineskip
        \ialign{$\m@th#1\hfil##\hfil$\crcr#2\crcr\sim\crcr } }}
\catcode`\@=12 

\def\lappeq{\lsim}
\def\gappeq{\gsim}

\def\slashb#1{\setbox0=\hbox{$#1$}#1\hskip-\wd0\dimen0=5pt\advance
       \dimen0 by-\ht0\advance\dimen0 by\dp0\lower0.5\dimen0\hbox
         to\wd0{\hss\sl/\/\hss}}

\def\defeq{\equiv}

\newcommand{\eqref}[1]{(\ref{#1})}   
\def\Toprel#1\over#2{\mathrel{\mathop{#2}\limits^{#1}}}

\begin{document}
\topmargin -0.5cm
\oddsidemargin -0.8cm
\evensidemargin -0.8cm

\pagestyle{empty}
\begin{flushright}
{CERN-TH/96-10}
\end{flushright}
\vspace*{5mm}
\begin{center}
{\bf ASTROPARTICLE PHYSICS - A Personal Outlook} \\
\vspace*{0.5cm}
{\bf John Ellis} \\
\vspace{0.1cm}
Theoretical Physics Division, CERN \\
CH - 1211 Geneva 23 \\
\vspace*{2cm}
{\bf ABSTRACT} \\
\end{center}
\vspace*{3mm}

At the request of the organizers, this talk surveys some of
the hot topics discussed at this meeting, giving my {\it subjective
views} on them. Subjects covered include the present age and
Hubble expansion rate of the Universe - {\it inflation theorists need
not yet abandon $\Omega = 1$}, theories of structure formation
in the light of COBE and other data - {\it my favourite is a
flat spectrum of initial perturbations subsequently amplified
by mixed hot and cold dark matter}, neutrino masses and
oscillations - {\it the only experimental indication I take
seriously at the moment is the persistent solar neutrino
deficit}, the lightest supersymmetric particle - {\it which
may behave differently if conventional assumptions are relaxed},
and the axion - {\it much of the window between limits from
SN 1987a and cosmology will be explored in an ongoing experiment}.
Finally, I present a chronology of some possible interesting future
experiments.

\vspace*{3cm}

\noindent
\rule[.1in]{16.5cm}{.002in}

\noindent
$^{*)}$ Invited talk presented at the IVth Workshop on
Theoretical and Phenomenological Aspects of Underground Physics
(TAUP 95), Toledo, Spain, September 1995 - to appear in the
Proceedings.
\vspace*{0.5cm}

\begin{flushleft} CERN-TH/96-10 \\
January 1996
\end{flushleft}

\vfill\eject
\pagestyle{plain}

\section{Introduction}

    This lecture is concerned with the three questions raised by
the Mayor of Toledo at the reception held during this conference:

{\bf ?`De d\'onde venimos?} Namely, what is the origin of the structure we
see in the Universe?

{\bf ?`Qu\'e somos?}
Namely, what is the nature of the Dark Matter around
us? and

{\bf ?`Ad\'onde vamos?} Namely, what lies beyond the Standard Model?

\section{On the Origin of Structure in the Universe}

\subsection{How Much Dark Matter?}

    Naturalness and inflation \cite{infl}
suggest that the density averaged over the
universe as a whole should be very close to the critical density, which
marks the boundary between a universe that expands forever and one which
eventually collapses, i.e. $\Omega\defeq\rho/\rho_c\simeq$1.  On the
other hand, the matter we can see shining in stars, in dust, etc.
amounts only to $\Omega \simeq 0.003$ to $0.01$ \cite{copi},
as seen in fig.~1.

    The commonly-agreed concordance
between big bang nucleosynthesis calculations \cite{Copietal}
and the observed
abundances of light elements suggests that $\Omega_{baryons}
\lappeq 0.1$, as also seen in fig.~1. This concordance has
recently been questioned, and it has even been suggested that big
bang nucleosynthesis may be in crisis \cite{Hata}. I do not share
this view (see also \cite{OS},\cite{Copietal}\cite{OlSc}).
For one thing, I have long believed that all the systematic errors
in the relevant physical quantities were not taken fully into
account \cite{eens},\cite{subir},\cite{highz}.
For another, I have less
 faith than some
\cite{Hata} in models of the chemical evolution of the galaxy
(see also \cite{Copietal,OlSc}). Finally, I would prefer not
to treat systematic errors as ``top hats", as was done in \cite{Hata},
which cuts off the tails of the distributions, and leads to estimates
of confidence levels that are difficult to interpret.

    Also shown in fig.~1 is an estimate of $\Omega_{baryons}$
from X-ray observations of rich clusters
\cite{Xray}, which tends to lie
somewhat higher than the big bang nucleosynthesis estimate.
However, the original rich cluster estimate was made in a pure
cold dark matter model. It is modified in the type of mixed
dark matter model to be discussed later \cite{newrich}, and could
also be reduced if the clusters are not virialized. In any
case, the possible discrepancy in fig.~1 is not very significant
for values of $H_0$ in much of the favoured range discussed below.

    The big bang nucleosynthesis estimate of $\Omega_{baryons}$ is
comparable  to the amount $\Omega_{halo}$ of matter that is
suggested by observations of rotation curves \cite{rotcurv} and the
virial theorem to be contained in galactic haloes: $\Omega_{halo}
\simeq 0.1$. Mathematically, the galactic haloes could in
principle be purely
baryonic, although they seem unlikely to be made out of gas, dust or
``snow balls" \cite{hegyi}. As you know,
there has recently been considerable interest
in the possibility that haloes might be largely composed of ``brown
dwarfs" weighing less than $\simeq 1/10$ of the solar mass.
The searches for such ``failed stars" in our galactic halo
via microlensing \cite{Pacz}
of stars in the Large Magellanic Cloud
in fact indicate \cite{lmc} that only a fraction
\beq
f = 0.20^{+0.33}_{-0.14}\,{\rm [MACHO]},\, <0.5\,{\rm [EROS]}
\label{E71}
\eeq
is composed of brown dwarfs \cite{Masso},\cite{newMACHO},
assuming a simple spherical halo model,
which would have a local density
\beq
\rho_{halo} = 0.3 \ \hbox{GeV/cm}^3 \times 1.5^{0\pm1}
\label{E72}
\eeq
The possibility has recently been reconsidered \cite{flat}
that our halo
is in fact significantly flattened, in which case the estimate
(\ref{E72}) of the local density should be increased, and the
brown dwarf fraction (\ref{E71}) correspondingly decreased.
To be on the conservative side when discussing cold dark matter
detection rates later in this talk, I will retain the spherical
halo estimate (\ref{E72}).

It should be emphasized that, in the standard theory of structure
formation reviewed in the next section, our halo {\it must} contain
a large fraction of non-baryonic cold dark matter
\cite{GatTurn}. On the other hand,
conventional infall models of galaxy formation suggest \cite{EllSik}
that our halo is unlikely to be composed mainly
of massive neutrinos, at least if their mass is chosen to yield
$\Omega_{hot} \simeq 0.2$ as suggested in the next section.
These observations follow from the need
for cold dark matter to boost galaxy formation, whereas
the phase-space density of neutrinos is
severely restricted \cite{TremaineGunn}.
The dominant component of our galactic halo
(\ref{E72}) should therefore be some form of cold dark matter.

    Before addressing in more detail the nature of the
non-baryonic dark matter,
I will first comment on the age and Hubble expansion rate of the
Universe, which have recently been the subject of some controversy
\cite{contro}.  Globular clusters
seem to be at least $14 \pm 3$ Gyr old, and nucleocosmochronology
suggests an age of $13 \pm 3$ Gyr \cite{copi}. Taken together,
these constraints suggest that the Universe cannot be younger
than $10$ Gyr, and that a greater age would be more comfortable.
The question is whether such an
age is compatible with current estimates of the Hubble constant
$H_0$ km/sec/Mpc, some of which are listed in Table~1. These
may be combined \cite{mrr} to yield the estimate
\beq
H_0 = 66 \pm 13
\label{hzero}
\eeq
where the central value is statistical, and the error is
supposed to be realistic, particularly in view of the fact that
any determination of $H_0$ involves
the combination of many steps. For example,
there has recently been a second determination based on
Hubble Space Telescope observations \cite{Leo}
of Cepheid variables (which have their own
intrinsic uncertainties), which must rely on other rungs in the
cosmic distance ladder, such as the distance to the Large
Magellanic Cloud, as well the extrapolation from Leo to the
Coma cluster. Errors in all of these must be combined in
order to arrive at the total uncertainty in $H_0$.
 \begin{center}
 \[ \left.  \begin{array}{l}  55  \pm 8 \\ 67 \pm 7 \end{array}~ \right\}
{\rm Type ~IA~ supernovae} \begin{array}{l} ~~{\rm (Sandage~ et~ al.)}
\\
~~{\rm (Riess ~et~
al.)} \end{array}\]
\[
 \begin{array}{l}  73  \pm 13  \end{array}~~
{\rm Type ~II~ supernovae} \begin{array}{l} ~~~~{\rm (Schmidt~ et~ al.)}
 \end{array}\]
\[ \left.  \begin{array}{l}  60  \pm 10 \\ 70 \pm 25 \end{array}
\right\}
{\rm Gravitational ~Lensing} \begin{array}{l} {\rm (Lehar~ et~ al.)} \\
{\rm (Wilkinson ~et~
al.)} \end{array}\]
\[
 \begin{array}{l}  55  \pm 17  \end{array}~~
{\rm Sunyaev-Zeldovich} \begin{array}{l} ~~{\rm (Birkinshaw~ et~ al.)}
 \end{array}\]
\[
 \begin{array}{l}  80  \pm 17  \end{array}~~
{\rm Virgo ~Cepheids} \begin{array}{l} ~~~~~~~~~~{\rm (Freedman~ et~
al.)}
 \end{array}\]
\[
 \begin{array}{l}  69  \pm 8  \end{array}~~
~~{\rm Leo ~I~ Cepheids} \begin{array}{l} ~~~~~~~~~~{\rm (Tanvir~ et~
al.)}
 \end{array}\] 

Table 1 - Recent determinations of $H_0$ (in km/s/Mpc)
\end{center}

The range (\ref{hzero})
is shown on the vertical axis
of fig.~1, where we see that there is no incompatibility between
the age of the Universe being $10$ Gyr old and
$\Omega = 1$ as wanted by inflation \cite{infl},
as long as $H_0$ is in the lower part of the range (\ref{hzero}).
Therefore, I see no immediate
need for inflation theorists to explore
models in which $\Omega$ is significantly below unity \cite{lowO},
which do not, in any case, look very natural to me.
Assuming that indeed $\Omega \simeq 1$,
at least $90 \%$ of the matter in the Universe must be unseen
non-baryonic dark matter.

\subsection{Hot or Cold Dark Matter?}

In addition to the above arguments based on contributions to
$\Omega$, non-baryonic dark matter is required for structure
formation, because it enables density perturbations to grow
via gravitational instability even before recombination,
while perturbations in the conventional baryonic matter density
are still restrained by the coupling to radiation. Which
structures form when depends
whether the non-baryonic dark matter was relativistic or
non-relativistic at the cosmological epoch when structures such as
galaxies and clusters began to form, which is the distinction
between ``hot" and ``cold".  Whether you favour hot or cold
dark matter depends on your favourite theory of structure formation.
If you believe that its origins lie in an approximately scale-invariant
Gaussian random field of
density perturbations, as suggested by inflationary models
\cite{perts}, then
you should favour cold dark matter.  This is because it enables
perturbations to grow on all distance scales, whereas relativistic
hot dark matter escapes from small-scale perturbations, whose growth
via gravitational instabilities is thereby stunted
\cite{sformation}.  Thus galaxies
form later in a scenario based on Gaussian fluctuations and hot dark
matter than they would in a scenario with cold dark matter,
indeed, too late. For this
reason, the combination of Gaussian perturbations with
cold dark matter has come to be regarded as the ``standard model"
of structure formation.  However, if you believe that structures
originated from seeds such as cosmic strings
\cite{strings}, then you should prefer
hot dark matter, because cold dark matter would then give too much
power in perturbations on small distance scales.

    Fig. 2  shows a compilation \cite{comp}
of data on the power spectrum of
astrophysical perturbations,
as obtained from earlier COBE
\cite{COBE} and other observations
of the cosmic microwave background
radiation, and direct astronomical
observations of
 galaxies and clusters. Subsequent to this compilation,
data from the full 4 years of COBE DMR data have been made available
\cite{COBE4}. These show no indications of non-Gaussian correlations
\cite{COBEGau}, and are consistent with a scale-invariant spectrum
\cite{COBEflat}, in agreement with inflationary models. However,
models of structure formation based on
cosmic strings would not
predict non-Gaussian correlations observable in the present data,
and would also yield a flat spectrum. Therefore, such models
cannot yet be excluded, though I will not address them further
in this talk \cite{strings}.

The overall normalization of
the perturbation spectrum is of interest to inflation theorists,
since it specifies the scale of the inflationary potential in
field-theoretical models. Parametrizing this by
$V = \mu^4 \bar V$, where $\bar V$ is a dimensionless function
of order unity, one finds that
\beq
\delta \rho / \rho \simeq \mu^2 G_N
\label{delrho}
\eeq
Taking the normalization of $\delta \rho / \rho$ from the
COBE data \cite{COBE}, one may estimate
\beq
\mu \simeq 10^{16} GeV
\label{inflscale}
\eeq
which is eerily close to the usual estimate
\cite{susygut} of the scale
of supersymmetric grand unification. A related quantity of
physical interest is the mass of the quantum of the inflationary
field, the inflaton:
\beq
m_{infl} \simeq 10^{13} GeV
\label{inflmass}
\eeq
which may have implications for baryogenesis and
neutrino masses, as discussed later. The scale (\ref{inflscale})
also determines the reheating
temperature at the end of the inflationary epoch, which is
of relevance to calculations of the potentially-dengerous
relic gravitino abundance \cite{gravitino} in supersymmetric
models.

The perturbations dicovered by COBE and experiments may in
general be a combination of density (scalar)
and gravity wave (tensor) fluctuations $A_{S,G}$, whose ratio
depends on details of the inflationary potential:
\beq
A_S / A_G = \sqrt{4 \pi G_N} H / |H'|
\label{ratio}
\eeq
The ratio (\ref{ratio}) exceeds unity if the inflaton field
accelerates during inflation, as expected, but the COBE
experiment is sensitive to the combination
$\simeq 25~A_G^2 /2~A_S^2$,
so gravity waves could be important. Nevertheless, one usually
assumes, as above,
 that scalar perturbations are dominant. A goal for
future experiments is to disentangle the scalar and tensor
contributions, and to measure the possible `tilts' of their
spectra:
\beq
n_{S,G} \defeq 1\,-\,(d/d\,ln\,\lambda) A_{S,G}
\label{tilt}
\eeq
so as to map out the inflaton potential \cite{map}.

Fig.~3 compiles data on fluctuations in the cosmic microwave background
radiation \cite{Tegmark}, and
provides the basis for a discussion of the issues arising in these future measurements. The original COBE
measurements at scales larger than
 the horizon at
recombination are conventionally
interpreted as due to the Sachs-Wolfe effect:
\beq
\delta T / T \simeq - \delta \phi / 3
\label{SachsWolfe}
\eeq
where $\phi$ is the gravitational potential.
There are by now many detections in the region
within the horizon at recombination, where the first
D\"oppler peak is expected to appear, with
\beq
\delta T / T \simeq v,
\label{Doppler}
\eeq
where $v$ is the baryonic matter velocity.
The existence of this first D\"oppler peak
cannot yet be
regarded as confirmed, but the outlook for
models which do not predict it
does not look very bright \cite{KogMin}.
The COBE data alone yield an
error of $\pm0.3$ on the effective spectral index \cite{COBEflat},
whereas a combined fit to the available
data indicates \cite{fit} the following range:
\beq
n \simeq 1.1 \pm 0.1
\label{index}
\eeq
Cosmic string models \cite{strings} are consistent with this and the
apparently Gaussian nature of the fluctuations seen at large
scales: a key test will be whether they still look Gaussian
in the region of the first D\"oppler peak. Within the standard
model of structure formation, the height of the this
peak will be a measure of $\Omega_{baryons}$, and its location
on the horizontal axis is sensitive to the total $\Omega$: $l\sim 220/\sqrt{\Omega}$.
Cold dark matter and related models predict further D\"oppler
peaks, which can only be resolved with a higher-resolution
experiment. Recall that the COBE resolution of a few degrees
includes a comoving volume that will later contain
several hundred clusters of galaxies: future experiments are
aiming at resolutions of a fraction of a degree. The drop-off
in fig.~3 visible
at still smaller scales is due to the thickness of the last
scattering surface.

The solid line which does not
quite pass through all the points
in fig.~2 is one calculated in the
above-mentioned standard model of Gaussian scalar fluctuations and
cold dark matter, assuming there is no tilt in the initial spectrum.
Crucial tests of this and other models of structure formation
will be provided by future measurements of the cosmic microwave
background radiation and of larger-scale structure in the region
of the bump in fig.~2, e.g., by the proposed
COBRAS/SAMBA satellite and the ongoing Sloan Digital Sky Survey.
The present discrepancies from this curve indicate that there is less
perturbation power at small distances than would be expected in this
theory, compared with the COBE normalization at large distance scales.

    This and other observations have suggested that it may be necessary
to modify the pure cold dark matter model.  Several suggestions have
been offered, including a non-zero cosmological constant and a tilt
in the spectrum of Gaussian perturbations away from scale
invariance.  However, the preferred scenario seems to be an admixture
of hot dark matter together with the cold, resulting in the following
cocktail recipe for the Universe \cite{mdm}:
\beq
\Omega_{cold} \simeq 0.7\,,~~     \Omega_{hot} \simeq 0.2,\,
~~     \Omega_{baryons} \lsim 0.1
\label{E73}
\eeq
The way in which this scenario works
is illustrated in fig.~4.  Hot dark
matter alone would give a spectrum of perturbations that dies out
at small scales, whereas hot dark matter does not.  Combining the two,
one can reconcile the relatively high COBE normalisation at large scales
with the relatively small perturbations seen at small scales.

    Fig.~5, which is adapted from \cite{Caldwell},
illustrates the performances of various dark matter
models of structure formation, as compared to measurements on
various different distance scales. We see that a pure cold dark
matter model has severe problems at smaller scales. These may be
somewhat alleviated by the introduction of biasing or a cosmological
constant $\Lambda$, but there are still problems on galactic
scales. A mixed dark matter model ({\ref{E73}) with the hot
dark matter provided by a single neutrino species of mass
$\simeq 5$ eV works quite well, except possibly for the density
of clusters. The authors of \cite{Caldwell} prefer for this
reason a model with two neutrino species each weighing
$\simeq 2.5$ eV, which is also
motivated by their interpretation of the LSND
experiment \cite{LSND}. However, I do not see the necessity for this
embellishment of the mixed dark matter model, and remain to be
convinced by the LSND data, as I discuss in the next section.

\section{On the Nature of the Dark Matter}

\subsection{Neutrino Masses and Oscillations}

    Theorists have been saying for years that there is no
fundamental reason why neutrino masses should vanish, and
oscillations are inevitable if they are non-zero. However,
for the time being, we only have the following
upper limits on neutrino masses:
\beq
m_{\nu_e} < 4.5\,\hbox{eV},\,m_{\nu_{\mu}} < 160\,\hbox{KeV},\,
m_{\nu_{\tau}} < 23\,\hbox{MeV}
\label{numasses}
\eeq
Cosmology in the form of big bang nucleosynthesis is close to
strengthening the above upper limit on $m_{\nu_{\tau}}$ to a
fraction of an MeV \cite{nutaumass}, if the $\nu_{\tau}$ is a
long-lived Majorana particle.

    There are many models for neutrino masses \cite{Valle},
which I will not discuss here. Instead, I will be inspired by
the simplest see-saw mass matrix \cite{seesaw}:
\beq
(\nu_L, \bar\nu_R) \left(\matrix{m_M&m_D\cr m_D & M_M}\right)~~
\left(\matrix{\nu_L\cr \bar\nu_R}\right)
\label{seesaw}
\eeq
where $m_M$ and $M_M$ are Majorana masses for the left- and
right-handed neutrinos $\nu_{L,R}$, respectively, and $m_D$
is a Dirac mass coupling $\nu_L$ and $\nu_R$. All of
$m_{M,D},M_M$ are to be understood as matrices in flavour space.
We expect $M_M$ to be comparable (on a logarithmic scale) with
the grand unification scale $M_X$, and the Dirac masses $m_D$ to
be comparable with the corresponding charge 2/3 quark masses
$m_{2/3}$.
We know from the experimental absence to date of neutrinoless
double-$\beta$ decay that
\beq
<m_{\nu_e}>_M~\lappeq~1/2~\hbox{eV}
\label{double}
\eeq
and diagonalization of (\ref{seesaw}) naturally suggests
that
\beq
m_{\nu}~\simeq~m_D^2/M_M
\label{small}
\eeq
for the known light neutrinos. If indeed $m_D \simeq m_{2/3}$,
we may expect that for the three light neutrino flavours
\beq
m_{\nu_i} \simeq {m_{{2/3}_i}^2 \over M_{M_i}}
\label{nuhier}
\eeq
The heavy Majorana masses $M_{M_i}$ are not necessarily universal,
but (\ref{nuhier}) nevertheless suggests that
\beq
m_{\nu_e} << m_{\nu_{\mu}} << m_{\nu_{\tau}}
\label{much}
\eeq
and the $\nu_{\tau}$ mass could be in the range of interest
to hot or mixed dark matter models if $M_{M_3} \simeq 10^{12}$
GeV.

To my mind, the most serious evidence for neutrino oscillations,
and, by extension, for neutrino masses, is the persistent and
recurrent solar neutrino deficit seen by four experiments
\cite{solarnu}, as compared with standard solar model
calculations \cite{Bahcall}. There
was much heated debate at this meeting on the interpretation of
these data \cite{debate}, and on the uncertainties in the
theoretical calculations. It seems to me that the crispest way
of posing the
dilemma is to plot the data in more than one dimension, for
example in the two-dimensional representation of fig.~6
\cite{nuplane}. Taken
at face value, all
the experiments indicate a strong suppression
of the Beryllium neutrinos, and a weaker suppression of the
Boron neutrinos, compared with the predictions of \cite{Bahcall}.
This major feature is very difficult to explain in plausible
modifications of this standard solar model, which tend to
suppress the Boron neutrinos more than the Beryllium neutrinos,
as also seen in fig.~6 \cite{nuplane}.
Note in particular the dotted curve
which corresponds a simply changing the central temperature
of the Sun, as happens in low-opacity
and simple mixing models. These cannot
explain the data, if the latter are taken at face value.
Some models which use different input nuclear cross sections
\cite{DS} even fall below the low-temperature curve:
they may explain the Boron deficit, but {\it a fortiori} they
cannot explain
simultaneously the Beryllium deficit, as seen clearly in this
two-dimensional plot.

    As we heard at this meeting \cite{helios},
the helioseismologists are
now making it very difficult even to reduce the central
temperature of the Sun. They are able to verify that the sound
speed, which is closely related to the temperature, agrees
with the standard solar model to within about $1 \%$ down to
about $5 \%$ of the solar radius. Indeed, the helioseismological
talk here \cite{helios}
included revised estimates of the Boron neutrino
flux that were {\it even higher} than the standard solar model.
One other possibility that was mentioned at this meeting was
that slow convection could alter significantly the standard
solar model predictions \cite{haxton},
but this remains to be demonstrated.

Two major new solar neutrino experiments will start taking
data next year, namely Superkamiokande \cite{SK} and SNO
\cite{SNO}. The former will
provide mind-boggling statistics for solar neutrinos, and for
any supernova explosion inside our galaxy. The SNO experiment
should be able to tell us whether the Beryllium neutrinos are
really absent, or ``merely" converted into another $\nu$
species, thanks to its aim of measuring both the charged- and
neutral-current reactions. Many other solar neutrino experiments
are on the drawing boards, including Borexino \cite{Borexino}
which is also aimed at the Beryllium neutrinos, as well as
HELLAZ \cite{HELLAZ} and HERON \cite{HERON}.

In my view, there {\it is} a solar neutrino problem, and novel
neutrino physics is the most likely explanation, though certainly
not yet established. The above-mentioned experiments may resolve
the issue. To make a useful theoretical contribution to the
continuing debate, I believe it is insufficient to point to
some possible modification of the standard solar model without
giving good reason to think that (a) it is consistent with {\it all}
the constraints, including, e.g., those from helioseismology,
and (b) it modifies the standard solar model predictions outside
the quoted errors. To support any such claim, a valuable
contribution should include some solid calculations.

If one does indeed take the solar neutrino deficit as an
indication for novel neutrino physics, I am inclined to plump
for mass mixing and oscillations rather than magnetic effects
\cite{Okunetal}
(the required magnetic dipole moment seems very high, and I am
not impressed by claims of a time dependence in the Homestake
data), and prefer the MSW scenario to vacuum oscillations.
In view of the prejudice (\ref{much}), the most likely
interpretation then becomes $\nu_e \rightarrow \nu_{\mu}$
oscillations, with
\beq
\Delta m^2 \simeq 10^{-5}\,\hbox{eV}^2\,\hbox{and}
\,sin^2\theta \simeq 10^{-2}
\label{MSW}
\eeq
provided by
\beq
m_{\nu_e} << m_{\nu_{\mu}} \simeq 10^{-3}\,\hbox{eV}
\label{MSW2}
\eeq
Scaling $m_{\nu_{\mu}}$ (\ref{MSW2}) up by $m_t^2/m_c^2$
(\ref{nuhier}), it is easy to imagine that there could be
$\nu_{\tau}$ dark matter with a mass around $10$ eV. One of
the most appealing aspects of this scenario is that it may
soon be tested in accelerator $\nu_{\mu} \rightarrow \nu_{\tau}$
oscillation experiments, since it suggests that $\delta m^2
\simeq 100$ eV$^2$, and many models for $\nu$ masses
further suggest a
value of $sin^2 \theta$ in the range accessible to the CHORUS
\cite{CHORUS} and NOMAD \cite{NOMAD} experiments
that are already taking data at CERN, or
the planned COSMOS experiment \cite{COSMOS} at Fermilab. If
needed, increased sensitivity could in principle be attained
with a next-generation detector \cite{NAUSICAA}.

This interpretation of the solar neutrino data also offers
the possibility of an appealing scenario for cosmological
baryogenesis \cite{FY}. After an inflationary epoch at the scale
(\ref{inflscale}), the inflaton with mass (\ref{inflmass})
can decay into a massive $\nu_R$ state, which then decays
out of thermal euilibrium. Diagrams of the type shown in
fig.~7 may produce a lepton asymmetry \cite{FY},\cite{ELNO}
\beq
\epsilon = {1\over 2\pi (\lambda^+_L\lambda_L)_{ii}}
\sum_j ({\rm lm} \left[ (\lambda^+_L\lambda_L)_{ij}\right]^2)
f\left({M^2_j\over M^2_i}\right)
\label{las}
\eeq
where the $\lambda_L$ are Yukawa couplings, $(i,j)$ are
generation indices, and $f$ is a kinematic function of the
$\nu_{R_i}$ masses $M_i$. The asymmetry $\epsilon$
is subject to subsequent reprocessing by electroweak
sphalerons \cite{sphal}
to produce a baryon asymmetry. This scenario can
be valid only if $m_{infl} > m_{\nu_R}$, which imposes a
lower limit on the inflation scale (\ref{inflscale}) and/or
a lower limit on the light neutrino mass (\ref{nuhier}).

In the longer term, ideas are afoot for long baseline
$\nu$ oscillation experiments between Fermilab and Soudan II,
between KEK and Superkamiokande, and between CERN and either
the Gran Sasso laboratory and/or the NESTOR underwater detector
now under construction \cite{Cav}. In the case of the possible
CERN-based experiments, the $\nu$ beam would be produced by a $120$ GeV
proton beam extracted from the SPS and directed along the
planned transfer line to the LHC. This type
of experiment would address principally the question of
possible atmospheric $\nu$ oscillations raised by the Kamiokande
experiment \cite{atmo}. I am not yet convinced of the reality of this
effect, and would like to see it confirmed with convincing
statistics by a large experiment using a completely different
experimental approach, as well as in Superkamiokande. Just for
fun, let me mention the idea for
what would surely be the ultimate earth-based $\nu$
oscillation experiment, namely to send a beam from CERN or
Fermilab to Superkamiokande. This would involve digging a beam
and decay tunnel inclined downwards at some $40$ degrees, which
would certainly amuse the civil engineers!

There is one final possibility of novel neutrino physics that I
would like to address, namely the suggestion that neutrino
oscillations may provide an explanation of the LSND data
\cite{LSND}. as you
see in fig.~8, there is not much room for this explanation,
given the constraints imposed by other experiments. Here
again, I would like to see confirmation from a different
experiment, as well as more data from the LSND experiment
itself, particularly in view of the fact that
there is no consensus yet on its interpretation \cite{Hill}.
Fortunately, we may not have to wait long, as the LSND group
promises us more information in the near future
\cite{Caldwell}, and reactor
experiments should soon be able to explore the region of interest.

\subsection{Lightest Supersymmetric Particle}

    My favourite candidate for cold dark matter
\cite{ehnos}, the lightest
supersymmetric particle (LSP) is expected to be stable in many
models, and hence present in the Universe as a cosmological relic
from the Big Bang. This is because supersymmetric particles possess
a multiplicatively-conserved quantum number called $R$ parity
\cite{Fayet}, which takes
the values $+1$ for all conventional particles and $-1$ for all
their supersymmetric partners.  Its conservation is a consequence
of baryon and lepton number cancelation, since
\beq
R = (-1)^{3B + L + 2 S}
\label{E74}
\eeq

There are three important consequences of $R$ conservation:
\begin{enumerate}
\item  Sparticles should always be produced in pairs.
\item  Heavier sparticles should decay into lighter ones.
\item  The LSP should be stable, since it has no legal decay mode.
\end{enumerate}

    In order to avoid condensation
into galaxies, stars and planets such as ours, where
it could in principle be detected in searches for anomalous heavy
isotopes \cite{isotopes}, it was argued in
\cite{ehnos} that any supersymmetric relic LSP should be
electromagnetically neutral and possess only weak interactions.
Scandidates in the
future sparticle data book include the sneutrino $\tilde\nu$ of spin
$0$, some form of ``neutralino" of spin $1/2$, or the gravitino
$\tilde G$ of spin $3/2$.  The sneutrino is essentially excluded by
the LEP experiments which measured the decay of the $Z^0$ into
invisible particles, which have
counted the number of light neutrino species: $2.991 \pm  0.0016$
\cite{Renton}, which does not leave space for any
sneutrino species weighing less than
${1\over2}M_{Z}$, and by underground experiments to be discussed in the
next section, which exclude a large range of heavier sneutrino masses
\cite{xsneutrino}.
Since the gravitino is probably impossible to discover, and is anyway
theoretically disfavoured as the LSP, we concentrate on the
neutralino \cite{ehnos}.

    The neutralino $\chi$ is a mixture of the photino $\tilde \gamma$,
the two neutral higgsinos $\tilde H_{1,2}^0$
expected in the minimal supersymmetric
extension of the Standard Model, and the zino
$\tilde Z$.  This is characterized essentially by three parameters,
the unmixed gaugino $m_{1/2}$, the Higgs mixing parameter $\mu$, and
the ratio of Higgs vacuum expectation values $\tan\,\beta$.  The
phenomenology of the lightest neutralino is quite complicated in
general,
but simplifies in the limit $m_{1/2} \rightarrow 0$, where $\chi$ is
approximately a photino state \cite{Goldberg},
and in the limit $\mu\rightarrow 0$,
where it is approximately a higgsino state.  As seen in fig.~9,
experimental constraints from LEP and the Fermilab collider in fact
exclude these two extreme limits \cite{erz}, so that
\beq
m_\chi\gsim (10\ \,\hbox{to}\ \,20) \,\hbox{GeV}
\label{E77}
\eeq

 Fig. 9 also indicates that there are generic domains of parameter
space where the LSP may have an ``interesting" cosmological relic
density \cite{density}, namely
\beq
0.1 \lsim \Omega_\chi H_0^2 \lsim 1
\label{E78}
\eeq
for some suitable choice of supersymmetric model parameters.  Fig.~10
displays the calculated LSP density in a sampling of phenomenological
models \cite{BergGond},   \cite{sample}
where we see that an interesting cosmological density is quite
plausible for LSP masses
\beq
20\ \hbox{GeV} \lsim m_\chi \lsim 300 \ \hbox{GeV}
\label{E79}
\eeq

For simplicity, this and most cancellations have made the simplifying
assumption of universality in the spectrum of sparticles, and have
also assumed that CP violation in the LSP couplings can be
neglected. Studies exploring the relaxation of these assumptions
have appeared recently. As seen in fig.~11a, it is much easier for
the LSP to be a higgsino-like state if the universality assumption
is relaxed \cite{Pok},\cite{Bott}, and, as seen in fig.~11b,
CP violation can relax the upper limit on the LSP mass \cite{FOS}.

 \section{On Searches for Cold Dark Matter Particles}

    In this section we first review some of the strategies that
have been proposed to search for relic neutralinos,
and then discuss cosmological axions. In considering interaction
rates for any given relic $\chi$, one should keep in mind the
correlation between the overall cosmological density $\Omega_{\chi}$
and the local halo density $\rho_{\chi}$. The most reasonable
assumption is that
\beq
\rho_{\chi} = (\Omega_{\chi}/\Omega_{cold})(1-f)\rho_{halo}
\label{reasonable}
\eeq
where $f, \rho_{halo}$ and $\Omega_{cold}$ are taken from
equations (\ref{E71}, \ref{E72}) and (\ref{E73}). It is not in general
worthwhile calculating rates for detection rates for relics
with uninteresting cosmological densities $\Omega_{\chi} << 1$,
and certainly not if one assumes the local density to be
$\rho_{halo}$.

\subsection{Annihilation in the Galactic Halo}

The first neutralino search strategy that we discuss is that
for the products of their annihilations
in our galactic halo \cite{haloann}.
Here the idea is that two self-conjugate $\chi$
particles may find each other while circulating in the halo, and have
a one-night stand and annihilate each other:
$\chi\chi\rightarrow\ell\ell,\bar q q$,
leading to a flux of stable particles such as $\bar p, e^+,\gamma,\nu$
in the cosmic rays.  Several experiments have searched for cosmic-ray
antiprotons \cite{previous},
with the results shown in fig.~12.
At low energies there
are only upper limits, but there are several positive detections at
higher energies, which are comparable with the flux expected from
secondary production by primary matter cosmic rays
\cite{Gaisser}. As also seen
in fig.~12, relic LSP annihilation in our galactic halo might produce
an observable flux of low-energy cosmic ray antiprotons somewhat below
the present experimental upper limits \cite{Freese}. This calculation
was made fixing the supersymmetric model parameters so that
$\Omega_{\chi} = 1$, and assuming that the local halo density
(\ref{E72}) is dominated by neutralinos $\chi$, and is subject to
uncertainties associated with the length of time that the $\bar p$'s
spend in our galactic halo.

Fluxes higher than those in \cite{Freese}
may be obtained if one considers neutralinos with $\Omega_{\chi}
< 1$, because they have larger annihilation cross sections. This is
a logical possibility, though it would mean that neutralinos would
not be the only (or even dominant) cold dark matter component, if
one retains the cocktail recipe (\ref{E73}). This would have the
corollary that the assumed local density should be
correspondingly reduced to some fraction of (\ref{E72}),
with a quadratic effect on the $\bar p$
flux, which is proportional to $\rho_{halo}^2$. In any case, it is
mathematically
impossible for the halo density (\ref{E72}) to be saturated by
neutralinos if the annihilation cross section is so large that
$\Omega_{\chi} < \Omega_{halo} \simeq 0.1$.
As already mentioned, in my view one
should be careful when quoting rates and limits on neutralino
parameters to check consistency with reasonable postulates on
$\Omega_{\chi}$ and $\Omega_{halo}$.

The flux estimates of \cite{Freese} may be interpreted as
suggesting that
\beq
\rho_\chi \lsim 10\,\rho_{halo}
\label{E80}
\eeq
and I do not believe it is possible to be much more precise at
the present time.
NASA and the DOE have recently approved a satellite experiment called
AMS \cite{AMS},
which should be able to improve significantly the present upper
limits on low-energy antiprotons, and may be able to start constraining
significantly supersymmetric models.

Finally, I note that it is also possible to derive limits
on supersymmetric models from the present experimental measurements of
the cosmic-ray $e^+$ \cite{HEAT} and $\gamma$ fluxes \cite{Bergstrom},
but these are not yet very constraining.

\subsection{Annihilation in the Sun or Earth}

    A second LSP detection strategy is to look for $\chi\chi$
annihilation inside the Sun or Earth.  Here the idea is that a relic
LSP wandering through the halo may pass through the Sun or Earth
\cite{solar},
collide with some nucleus inside it, and thereby lose recoil energy.
This could convert it from a hyperbolic orbit into an elliptic one,
with a perihelion (or perigee) below the solar (or terrestrial) radius.
If so, the initial capture would be followed by repeated scattering
and energy loss, resulting in a quasi-isothermal distribution within the
Sun (or Earth).  The resulting LSP population would grow indefinitely,
\`a la Malthus, unless it were controlled either by emigration,  namely
evacuation from the surface, or by civil war, namely annihilation within
the Sun (or Earth).  Evaporation is negligible for $\chi$ particles
weighing more than a few GeV \cite{Gould},
so the only hope is annihilation.  The
neutrinos produced by any such annihilation events would escape from
the core, leading to a high-energy solar neutrino flux ($E_{\nu}
\gappeq 1$ GeV).  This could be detected either directly in an
underground experiment, or indirectly via a flux of upward-going muons
produced by neutrino collisions in the rock. (By the way, LSPs in
the core of the Sun do not affect significantly its temperature,
and hence have no impact on the low-energy solar neutrino problem.)

The high-energy solar neutrino flux produced in this way is given
approximately by the following general formula \cite{EFR}:
\bea
R_\nu = &2.7\times 10^{-2} \,f\left(m_\chi/m_p\right)\,
\left({\sigma (\chi p \rightarrow \chi p)
\over 10^{-40}\, \hbox{cm}^2}\right) \hfill
\nonumber \\
&
\left({\rho_\chi\over 0.3\,\hbox{GeV cm}^{-3}}\right)
\left({300 \ \hbox{km\,s}^{-1}\over \bar v_{\chi}}\right)
\times F_\nu \hfill
\label{E81}
\eea
assuming that proton targets dominate capture by the Sun.
Here $f$ is a kinematic function, $\sigma (\chi\,p \rightarrow
\chi\,p)$ is the elastic LSP-proton scattering cross section,
$\rho_{\chi}$ and $\bar v_{\chi}$ are the local density and mean
velocity
of the halo LSPs, and $F_\nu$ represents factors associated with
the neutrino interaction rate in the apparatus.

There is an analogous formula for the production
of upward-going muons originating from the collisions in rock of
high-energy solar neutrinos, and rates in a sampling of supersymmetric
models are shown in fig.~13. While some models
are
 already excluded by unsuccessful searches, most are not
\cite{sample}.
We see in fig.~14 that searches for solar signals usually
constrain models more than searches for terrestrial signals,
though this is not a model-independent fact
\cite{sample}. As seen in fig.~15, in the long run
it seems that a search for upward-going neutrino-induced muons
with a $1$ km$^2$ detector could almost certainly detect LSP
annihilation \cite{Halzen}, if most of the cold dark matter is indeed
composed of LSPs.

Before leaving this subject, I would like to recall that MSW
oscillations may also be important for high-energy solar
neutrinos, as seen in fig.~16 \cite{EFM}. Until the possible
neutrino mass and oscillation parameters are pinned down, this
introduces another ambiguity into the above analysis. For the
time being, it would be conservative to quote upper
limits on fluxes assuming that
the neutrinos arriving at the detector are those for which the
detector has the smallest efficiency.

\subsection{Dark Matter Search in the Laboratory}

    The third LSP search strategy is to look directly for
LSP scattering off nuclei in the laboratory \cite{GW}.
The typical recoil energy
\beq
\Delta E < m_\chi v^2 \simeq 10
\left({m_\chi\over 10\,\hbox{GeV}}\right)\,\hbox{keV}
\label{E85}
\eeq

deposited by elastic $\chi$-nucleus scattering would probably
lie in the range of $10$ to $100$ keV. Spin-dependent
interactions mediated by $Z^0$ or $\tilde q$ exchange are likely
to dominate for light nuclei
\cite{Flores}, whereas coherent spin-dependent
interactions mediated by $H$ and $\tilde q$ exchange are likely
to dominate scattering off heavy nuclei
\cite{Griest}. The spin-dependent interactions on individual nucleons
are controlled by the contributions of the different flavours $q$
of quark to the total nucleon spin, denoted by
$\Delta q$. These have now been determined by polarized
lepton-nucleon scattering experiments with an accuracy
sufficient for our purposes. Translating the $\Delta q$ into matrix
elements for interactions on nuclei depends on the contributions
of the different nucleon species
to the nuclear spin, which must be studied using the shell model
\cite{Flores} or some other theory of nuclear structure
\cite{nuclear}. The spin-independent
interactions on individual nucleons are related to the different
quark and gluon contributions to the nucleon mass, which is also
an interesting phenomenological issue related to the $\pi$-nucleon
$\sigma$-term \cite{sigma}. Again, the issue of nuclear
structure arises when one goes from the nucleon level to coherent
scattering off a nuclear target. It is in particular necessary
to understand the relevant nuclear form factors, which are expected
to exhibit zeroes at certain momentum transfers \cite{zeroes}.

We will not discuss here the details of such nuclear calculations,
but present in figs~17 and 10 the results of a sampling of
different supersymmetric models
\cite{sample}. We see in fig.~17 that the
spin-independent contribution tends to dominate
 over the
spin-dependent one in the case of Germanium, though this is not
universally true, and would not be the case for scattering off
Fluorine \cite{Flores}.
In fig.~10 we plot the scattering rates off $^{73}$Ge,
where we see that there are many models in which more than
$0.01$ events/kg/day are expected, which may be observable
\cite{BergGond}.
The direct search for cold dark matter scattering in the
laboratory may be a useful complement to the searches for
supersymmetry at accelerators \cite{Flores}.

Fig.~18 shows as solid lines
upper limits from searches for elastic scattering in the laboratory, for
spin-dependent rates:
 \bea
&&\sigma_p^{dep}\, \lappeq 0.3\,\hbox{pb}\;\hbox{for}
\nonumber \\
&&20\,\hbox{GeV}\, \lappeq\,m_{\chi}\,\lappeq\,300\,\hbox{GeV}
\label{spindep}
\eea
in part (a), and for spin-independent scattering rates:
\bea
&&\sigma_p^{ind}\,\lappeq\,3~10^{-5}\,\hbox{pb}\;\hbox{for}
\nonumber \\
&&20\,\hbox{GeV}\, \lappeq\,m_{\chi}\,\lappeq\,300\,\hbox{GeV}
\label{spinindep}
\eea
in part (b) \cite{survey}. Also shown in fig.~18 as dashed lines
are corresponding upper limits from indirect searches of the
type discussed in the previous subsection. These may appear more
stringent, but involve more uncertainties, as already discussed.
Please note also that limits may also be obtained from studies of
tracks in ancient Mica \cite{Mica}, and from searches for
inelastic excitations by relic particles \cite{inelastic}.

 It should also be emphasized that the significances and relative
importances of these different search strategies are sensitive
to the usual assumption of universality in the sparticle masses.
This point is made in fig.~19, whose panel (a) shows results in
a sampling of ``universal" models, whilst panel (b) shows what
happens in a sampling of ``non-universal" models \cite{Pok},
\cite{Bott}.
The latter have yet to be explored so systematically.

\subsection{Axions}

The axion \cite{axion} is my second-favourite candidate for the cold
dark matter. As you know, it was invented to guarantee conservation of
$P$ and $CP$ in the strong interactions. These would otherwise
be violated by the QCD $\theta$ parameter,
which is known experimentally to be smaller than about $10^{-9}$
\cite{RPP}. The $\theta$ parameter relaxes to zero in any extension of
the Standard Model which contains the axion, whose mass and
couplings to matter that are scaled inversely by the
axion decay constant $f_a$.
The fact that no axion has been seen in any accelerator experiment
tells us that
\beq
f_a \gsim 1\,\hbox{TeV}
\label{E88}
\eeq
and hence that any axion must be associated with physics beyond
the scale of the Standard Model.

Axions would have been produced in the early Universe in the
form of slow-moving coherent waves that could constitute
cold dark matter. The relic density of these waves has been
estimated as \cite{adensity}
\bea
\Omega_a \,\simeq\,&
\left({0.6\times 10^{-5}\,\hbox{eV}\over m_a}\right)^{7/6}
\hfill
\nonumber \\
&\left({200\,\hbox{MeV}\over \Lambda_{QCD}}\right)^{3/4}
\left({75\over H_0}\right)^2
\label{E89}
\eea
which is less than unity if
\beq
f_a \lsim 10^{12}\,\hbox{GeV}
\label{E90}
\eeq
In addition to these coherent waves, there may also be axions
radiated from cosmic strings
\cite{radiation}, which would also be non-relativistic
by now, and hence contribute to the relic axion density and
strengthen the limit in equation (\ref{E90}).

The fact that the Sun shines photons rather than axions, or, more
accurately but less picturesquely, that the standard solar model
describes most data, implies the lower limit
\beq
f_a \gsim 10^{7}\,\hbox{GeV}
\label{E91}
\eeq
This has been strengthened somewhat by unsuccessful searches for
the axio-electric effect \cite{Sun},
in which an axion ionizes an atom. More
stringent lower bounds on $f_a$ are provided by the agreements
between theories of Red Giant and White Dwarf stars with the
observations \cite{stars}:
\beq
f_a \gsim 10^{9}\,\hbox{GeV}
\label{E92}
\eeq
Between equations (\ref{E90}) and (\ref{E92})
there is an open window in which the
axion could provide a relic density of interest to astrophysicists
and cosmologists.

Part of this window is closed by the observations of the
supernova SN1987a. According to the standard
theory of supernova collapse to form a neutron star,
$99 \%$ of the binding energy released in the collapse to the
neutron star escapes as neutrinos. This theory agrees \cite{neutrinos}
with the observations of SN1987a made by the Kamiokande \cite{Kam}
and IMB experiments \cite{IMB}, which means that most of
the energy could not have been carried off by other invisible
particles such as axions.

Since the axion is a light pseudoscalar boson, its couplings to
nucleons are related by a generalized Goldberger-Treiman relation
to the corresponding axial-current matrix elements, and these are
in turn determined by the corresponding $\Delta q$
\cite{GTaxion}. Specifically,
we find for the axion couplings to individual nucleons that
\bea
C_{ap} = &\,2[{-} 2.76\, \Du - 1.13\, \Dd + 0.89\, \Ds
\nonumber \\
&-\cos 2 \beta \, (\Du - \Dd - \Ds) ]\,,  \hfill
\nonumber\\
\label{E93}\\
C_{an} = &\,2[{-} 2.76\, \Dd - 1.13\, \Du + 0.89\, \Ds
\nonumber \\
&-\cos 2 \beta \, (\Dd - \Du - \Ds) ] \hfill
\phantom{\,,}
\nonumber
\eea
Evaluating the $\Delta q$ at a momentum scale around $1$ GeV, as is
appropriate in the core of a neutron star, we estimate
\cite{BjSRalphas} that
\bea
C_{ap} &=& ({-}3.9 \pm 0.4) - (2.68 \pm 0.06)
\cos 2 \beta
\nonumber\\
\label{E94}\\
C_{an} &=& (0.19 \pm 0.4)\, + \,(2.35 \pm 0.06)
\cos 2 \beta
\phantom{,}
\nonumber
\eea
which are plotted in fig.~20.

As in the case of LSP scattering, the uncertainties associated with
polarized structure function measurements are by now considerably
smaller than other uncertainties, in this case particularly those
associated with the nuclear equation of state. A particular
focus of attention here has been the suggestion \cite{Raffelt2}
that nucleon spin fluctuations in the supernova core might
suppress substantially the axion emission rate. In fact, sum-rule
considerations \cite{Sigl} suggest that this suppression may be less
important than first thought, though there is some shift in the
open part of the axion window \cite{newsn}.

The good news here is that an experiment \cite{axexp} is underway which should
be able to detect halo axions if they live in at least part of
this window.

\section{Prospects}

We have every reason to think that the near future will be a
very exciting period for astroparticle physics. As seen in
Table 2, many experiments are underway, under construction, or
being actively planned, which will contribute to resolving the
fundamental issues in this field. On the side of astrophysics and
cosmology, we have every reason to hope that a verified ``Standard
Model" of structure formation will soon emerge, and that the
nature of the invisible $90 \%$ or more of the matter in the
Universe may soon be resolved. On the side of particle physics,
we have every reason to hope that the resolution of these
astrophysical and cosmological issues will take us beyond the
current Standard Model of particle physics, a strait-jacket from
which accelerators have not yet been able to extract us. As is
seen in Table 2, it may in fact be the next generation of accelerator
experiments that creates these twin revolutions in astroparticle
physics.

\section*{Acknowledgements}

I thank Angel Morales, Mercedes Fatas and members of the Organizing
Committee for creating such a stimulating meeting.

 \newpage
\begin{figure*}[H]
\epsfig{figure=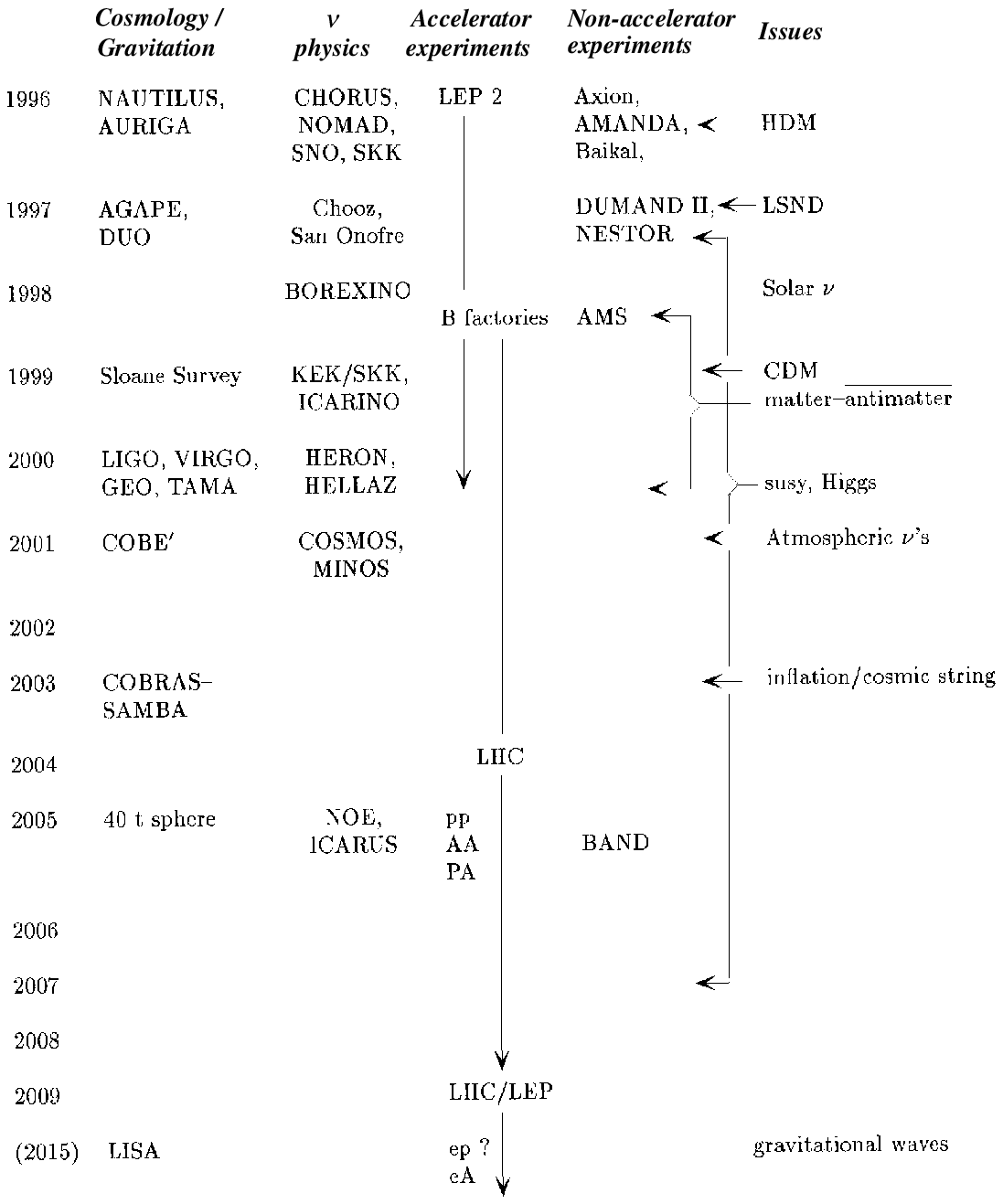,width=16cm}
\begin{center}Table~2 - Chronology of some possible future interesting
experiments and \\
the astroparticle physics issues they may resolve.
\end{center}
\end{figure*}

\vspace*{2cm}

\noindent
{\bf Figure Captions}
\begin{itemize}
\item[Fig. 1 - ]
The $(\Omega , H_0)$ plane, adapted from
\cite{copi}, which exhibits no serious
discrepancy between the average measured value of $H_0, \Omega = 1$, and
an age for the
Universe of 10$^{10}$ years. This plot also shows the estimates of the
present baryon
density $\Omega_{baryons}$ obtained from visible features in the
Universe, from Big Bang
Nucleosynthesis and from rich clusters. All the indications
are that
$\Omega_{baryons} \lappeq$ 0.1, so that at least 90\% of the matter in
the Universe is
non-baryonic dark matter.

\item[Fig. 2 - ]
A compilation of data on the primordial perturbation spectrum
\cite{comp}, compared with
a cold dark matter simulation assuming an initially scale-invariant
spectrum of Gaussian
fluctuations. The wave number $k \propto 1/d$,  where $d$ is
the distance scale of the perturbation.

\item[Fig. 3 - ]
Compilation of data on fluctuations in the cosmic microwave
background spectrum \cite{Tegmark}, which provides the basis for a
discussion of theoretical ideas and observational
issues in measuring fluctuations in the cosmic microwave
background radiation. The harmonic number $l \propto 1/d$.

\item[Fig. 4 - ]
Illustration how a mixed dark matter scenario \cite{mdm}
may reconcile the large-scale perturbations seen by COBE
\cite{COBE},\cite{COBE4}
with the relatively small magnitude of the
perturbations seen at small scales.

\item[Fig. 5 - ]
A sketch indicating the relative successes of different
models of structure formation, as compared with different types of
astrophysical and cosmological data.

\item[Fig. 6 - ]
A planar presentation \cite{nuplane}
of the solar neutrino deficits seen in
different experiments, compared with a selection of different
solar models.

\item[Fig. 7 - ]
Diagrams which may produce a lepton asymmetry in heavy neutrino
decay, which may subsequently be partly recycled by sphalerons into a
baryon asymmetry \cite{FY},\cite{ELNO}.

\item[Fig. 8 - ]
Confrontation of the LSND data \cite{LSND}
with constraints on neutrino
oscillation parameters provided by other experiments.

\item[Fig. 9 - ]
The cosmological relic density of the neutralino $\chi$
\cite{ehnos} may well (shaded
regions) lie in the range of interest to astrophysicists and
cosmologists \cite{density},
namely $0.1 \lappeq \Omega_\chi \lappeq 1$. 

\item[Fig. 10 - ]
Relic density of supersymmetric particles, calculated in a
sampling of different models \cite{BergGond}, together
with the estimated scattering rate on $^{76}$Ge.

\item[Fig. 11 - ]
Some results of relaxing the usual restrictive assumptions
made in analyses of supersymmetric relic particles: (a) the
supersymmetric relic is more likely to be a higgsino if scalar
masses are not universal \cite{Bott}, and (b) it may be heavier if there
are large CP-violating phases\cite{FOS}.

\item[Fig. 12 - ]
The results of experimental searches for cosmic-ray antiprotons
\cite{previous}, compared
with the  secondary fluxes expected from primary matter cosmic rays
\cite{Gaisser} and a supersymmetric model
\cite{Freese}.
The lower points with error bars are what should be
obtainable with the
AMS experiment \cite{AMS}.

\item[Fig. 13 - ]
The flux of upward-going muons expected from $\chi\chi$
annihilation inside the Sun in
a sampling of supersymmetric models \cite{sample}.

\item[Fig. 14 - ]
A comparison of the muon fluxes from the centre of the Sun and
Earth, as found in a
sampling of supersymmetric models \cite{sample}. 

\item[Fig. 15 - ]
A detector with an area of one square kilometer is likely
to able to detect upward-going muons from relic annihilations
\cite{Halzen}.

\item[Fig. 16 - ]
High-energy solar neutrinos produced by relic annihilations
inside the Sun may be sensitive to matter-enhanced oscillations
\cite{EFM}.

\item[Fig. 17 - ]
A comparison of the spin-dependent and spin-independent interaction
rates of relic
neutralinos $\chi$ with Germanium in a sampling of supersymmetric models
\cite{sample}.

\item[Fig. 18 - ]
A compilation of present upper limits on dark matter
scattering via (a) spin-dependent and (b)
spin-independent interactions
\cite{survey}.

\item[Fig. 19 - ]
The sensitivities of direct and indirect searches for
dark matter in (a) ``universal" models, and (b) ``non-universal"
models \cite{Bott}.

\item[Fig. 20 - ]
Axion couplings to the nucleon, as determined in \cite{BjSRalphas}
using polarized structure function data.

\end{itemize}

\end{document}